\documentclass[10pt,preprint]{aastex}
\usepackage{natbib}

\shorttitle{Synthetic Spacecraft Data}
\shortauthors{Klein, Howes, TenBarge, Bale, Chen, and Salem}

\newcommand\Alfven{Alfv\'en }
\newcommand\Alfvenic{Alfv\'enic }
\newcommand{\V}[1]{\mathbf{#1}} 
 
\newcommand{\zhat}{\mbox{$\hat{\mathbf{z}}$}} 
\newcommand{\xhat}{\mbox{$\hat{\mathbf{x}}$}} 
 
\newcommand{\appref}[1]{Appendix~\ref{#1}}
\newcommand{\figref}[1]{Figure~\ref{#1}}
\newcommand{\secref}[1]{\S\ref{#1}}
\newcommand{\eqref}[1]{equation~(\ref{#1})}

\usepackage{color}

\begin{document}


\title{Using Synthetic Spacecraft Data to Interpret \\Compressible 
Fluctuations in Solar Wind Turbulence}


\author{K.~G. Klein and  G.~G. Howes and J.~M. TenBarge}
\affil{Department of Physics and Astronomy, University of Iowa, Iowa City, 
Iowa 52242, USA.}

\and

\author{S.~D.~Bale\altaffilmark{1} and  C.~H.~K.~Chen  and C.~S.~Salem}
\affil{Space Science Laboratory, University of California, Berkeley, 
California 94720, USA.}

 \altaffiltext{1}{Department of Physics, University of California, Berkeley, 
California 94720, USA.}

\begin{abstract}
Kinetic plasma theory is used to generate synthetic spacecraft data to
analyze and interpret the compressible fluctuations in the inertial
range of solar wind turbulence. The kinetic counterparts of the three
familiar linear MHD wave modes---the fast, Alfv\'en, and slow
waves---are identified and the properties of the density-parallel
magnetic field correlation for these kinetic wave modes is
presented. The construction of synthetic spacecraft data, based on the
quasi-linear premise---that some characteristics of magnetized plasma
turbulence can be usefully modeled as a collection of randomly phased,
linear wave modes---is described in detail. Theoretical predictions of
the density-parallel magnetic field correlation based on MHD and
Vlasov-Maxwell linear eigenfunctions are presented and compared to the
observational determination of this correlation based on 10 years
of\emph{Wind} spacecraft data. It is demonstrated that MHD theory is
inadequate to describe the compressible turbulent fluctuations and
that the observed density-parallel magnetic field correlation is
consistent with a statistically negligible kinetic fast wave energy
contribution for the large sample used in this study. A model of the
solar wind inertial range fluctuations is proposed comprised of a
mixture of a critically balanced distribution of incompressible 
Alfv\'enic fluctuations and a critically balanced or more anisotropic
than critical balance distribution of compressible slow wave
fluctuations. These results imply that there is little or no transfer
of large scale turbulent energy through the inertial range down to
whistler waves at small scales.
\end{abstract}


\section{Introduction}
  Despite more than forty years of direct spacecraft measurements of
  turbulence in the near-Earth solar wind
\citep{Coleman:1968}, our understanding of turbulence in a magnetized 
plasma remains incomplete. One of the primary goals is to understand
the role of the turbulence in mediating the transfer of energy from
large to small scales. Within the turbulent inertial range of scales,
corresponding to spacecraft-frame frequencies of $10^{-4}$ Hz
$\lesssim f_{sc} \lesssim 1$ Hz or length scales $10^6$~km $\gtrsim
\lambda \gtrsim 10^2$~km, the fluctuations involved in this energy 
transfer are a mixture of compressible and incompressible fluctuations,
with around 90\% of the energy in the incompressible component
\citep{Tu:1995,Bruno:2005}. These incompressible fluctuations have 
been identifed as \Alfven waves \citep{Belcher:1971}, but the nature
of the compressible component remains uncertain.

The compressible turbulent fluctuations have often been interpreted
as a combination of magnetoacoustic (fast MHD) waves and
pressure-balanced structures \citep{Tu:1995,Bruno:2005}.  Early
studies of thermal and magnetic pressure fluctuations in the solar
wind found an anti-correlation of the thermal pressure and magnetic
pressure at timescales of 1~h, corresponding to an interval  of constant
total pressure, or a pressure-balanced structure (PBS)
\citep{Burlaga:1968,Burlaga:1970}. Further studies  found
evidence of PBSs out to 24~AU \citep{Burlaga:1990}. Related
investigations discovered a similar anti-correlation between the
density $n$ and magnetic field magnitude $B$ from 0.3~AU to 18~AU
on timescales ranging from several hours to 1.8  minutes
\citep{Vellante:1987,Roberts:1987b,Roberts:1987a,Roberts:1990}.
Theoretical studies of compressive MHD fluctuations in the low-Mach
number, high-$\beta$ limit interpreted these anti-correlated
density-magnetic field strength observations as nonpropagating
``pseudosound'' density fluctuations
\citep{Montgomery:1987,Matthaeus:1991}. Later, a more comprehensive observational 
investigation confirmed the general density-magnetic field magnitude
anti-correlation, but also identified a few positively correlated
intervals consistent with the magnetosonic (fast MHD) wave
\citep{Tu:1994}. Analysis of
\emph{Ulysses} observations found evidence for PBSs at inertial range
scales in the high latitude solar wind
\citep{McComas:1995,Reisenfeld:1999,Bavassano:2004}. Studies of the
electron density up to $f=2.5$~Hz also found pressure balanced
structures but interpreted these as ion acoustic (slow MHD) waves
and recognized that PBSs are simply the ion acoustic (slow MHD) wave
in the perpendicular wavevector limit \citep{Kellogg:2005}, a fact
previously noted by \citet{Tu:1994}. Recently, measurements of the
anti-correlation between electron density and magnetic field strength
indicated the existence of PBSs over timescales ranging from 1000~s
down to 10~s \citep{Yao:2011}.

An important consideration in the study of the compressible
fluctuations of solar wind turbulence is the fact that the mean free
path in the solar wind plasma is about 1~AU, so the dynamics over the
entire inertial range is weakly collisional. The implications of this
fact have not been seriously addressed in any of the aforementioned
studies of compressible fluctuations in solar wind turbulence.  The
MHD description is rigorously valid only in the limit of strong
collisionality, so a kinetic description is formally required to
describe the inertial range turbulence. In the limit $k_\parallel \ll
k_\perp$ predicted by anisotropic MHD turbulence
theories\footnote{Parallel and perpendicular are defined with respect
to the direction of the local mean magnetic field.}
\citep{Goldreich:1995,Boldyrev:2006}, it has been demonstrated that,
even in the weakly collisional limit, the turbulent dynamics of the
\Alfven waves decouples from the compressible fluctuations and 
is rigorously described by the equations of reduced MHD
\citep{Schekochihin:2009}.  The compressible fast and slow wave modes,
on the other hand, require a kinetic description to resolve both the
wave dynamics and the collisionless kinetic damping mechanisms. The
study presented here is the first to examine the properties of the
compressible fluctuations in the turbulent solar wind using
Vlasov-Maxwell kinetic theory. Specifically, we use the predicted
correlation between the density fluctuations and parallel magnetic
field fluctuations to determine the nature of the compressible
fluctuations in the solar wind.

In \secref{sec:collisionless}, we explore the connection between the
familiar linear wave modes in MHD and the corresponding kinetic wave
modes in Vlasov-Maxwell kinetic theory and demonstrate that weakly
collisional conditions do not change the qualitative properties of the
density-parallel magnetic field correlation. In \secref{sec:ssd}, we
discuss the quasi-linear premise upon which the method of synthetic
spacecraft data is based and describe in detail the procedure for
generating synthetic spacecraft data. The synthetic spacecraft data
predictions of the density-parallel magnetic field correlation based
on linear MHD and Vlasov-Maxwell eigenfunctions is presented in 
\secref{sec:ssd_predict}. A comparison of the synthetic spacecraft data
predictions to the observational determination of the density-parallel
magnetic field correlation is presented in \secref{sec:discuss},
showing a statistically negligible fast wave energy contribution to
the compressible fluctuations. The implications of this finding are
discussed before summarizing the findings of this investigation in
\secref{sec:conc}.

\section{Collisional vs.~Collisionless Dynamics of Compressible Fluctuations}
\label{sec:collisionless}


The inertial range of solar wind turbulence is observed to be a
mixture of incompressible and compressible motions, with around 90\%
of the energy due to the incompressible component
\citep{Tu:1995,Bruno:2005}.  If these fluctuations are interpreted as
some mixture of the three MHD linear wave modes, then \Alfven waves
are responsible for the incompressible component, while slow and fast
MHD waves make up the compressible component. In the
limit of large scales compared to the thermal ion Larmor radius, $k
\rho_i \ll 1$, these modes may be  distinguished by the correlation
between the density and parallel magnetic field fluctuations: fast
waves are positively correlated, slow waves are negatively correlated,
and the density and parallel magnetic field fluctuations are both zero
for
\Alfven waves. As the wave amplitude is increased to nonlinear levels,
even in the limit that they form discontinuities or shocks, these
qualitative properties persist, corresponding to tangential and
rotational discontinuities or fast and slow shocks
\citep{Baumjohann:1996}. In this section, we will explore the
properties of the kinetic counterparts to the fast and slow MHD wave
modes in the inertial range using Vlasov-Maxwell kinetic theory.

\subsection{ The Inertial Range Limit of the Compressible Linear Wave Modes}
\label{sec:model}

As derived in Appendix~\ref{app:MHDeig}, the normalized 
compressible MHD linear dispersion relation depends on only two parameters,
$\omega/(k v_A) =\overline{\omega}_{MHD}( \beta, \theta)$: the plasma
beta, $\beta = c_s^2/v_A^2$, where the sound speed is $ c_s = \sqrt{\gamma p_0/\rho_0}$
and the \Alfven velocity is $v_A = B_0/\sqrt{4 \pi \rho_0}$; and the angle
$\theta$ between the local mean magnetic field $\V{B}_0$ and the
direction of the wavevector. 

To establish precisely the connection between the three linear MHD
wave modes and their kinetic counterparts in the linear Vlasov-Maxwell
system, we specify a fully-ionized proton and electron plasma with
isotropic Maxwellian velocity distributions and a realistic mass ratio
$m_i/m_e=1836$. In general, the linear Vlasov-Maxwell dispersion
relation depends on five parameters: the ion plasma beta $\beta_i= 8
\pi n_i T_i/B_0^2$, which is the ratio of the ion thermal pressure to the
magnetic pressure\footnote{In this study, the Boltzmann constant is
absorbed into the temperature to yield temperature in units of energy.},
the normalized wavenumber $k \rho_i$, the angle $\theta$ between the
local mean magnetic field $\V{B}_0$ and the direction of the
wavevector, the ion to electron temperature ratio $T_i/T_e$, and the
ratio of ion thermal velocity to the speed of light $v_{ti}/c$. The
solution may then be expressed as $\omega=\omega_{VM} (\beta_i,k \rho_i,
\theta,T_i/T_e,v_{ti}/c)$ \citep{Stix:1992,Quataert:1998,Howes:2006}.
To connect to the single fluid theory of MHD, we take equal
ion and electron temperatures, $T_i/T_e=1$.  In the inertial range
limit, $k \rho_i \ll 1$, and for the non-relativistic conditions
appropriate to the solar wind, $v_{ti}/c \ll 1$, the normalized linear
Vlasov-Maxwell dispersion relation simplifies to $\omega/(kv_A)
=\overline{\omega}_{VM}( \beta_i, \theta)$. Since $\beta= \beta_i +
\beta_e = \beta_i(1+T_e/T_i)$, direct quantitative comparison 
between the solutions of the MHD and Vlasov-Maxwell linear dispersion
relations is possible by choosing $\beta= 2 \beta_i$ \citep{Howes:2009b}.

The Vlasov-Maxwell solutions presented in this section use the
parameters $k \rho_i =2.2 \times 10^{-4}$,
$\frac{v_{t_i}}{c}=10^{-4}$, and $T_i/T_e=1$, and employ a realistic
mass ratio $m_i/m_e=1836$. The complex eigenfrequencies are solved
numerically \citep{Quataert:1998,Howes:2006} with Bessel function sums
evaluated to 100~terms to ensure accurate results.


\subsection{Connection between MHD and Vlasov-Maxwell Linear Wave Modes}
\label{sec:connect}

\begin{figure}[p] 
\includegraphics[scale=1.0]{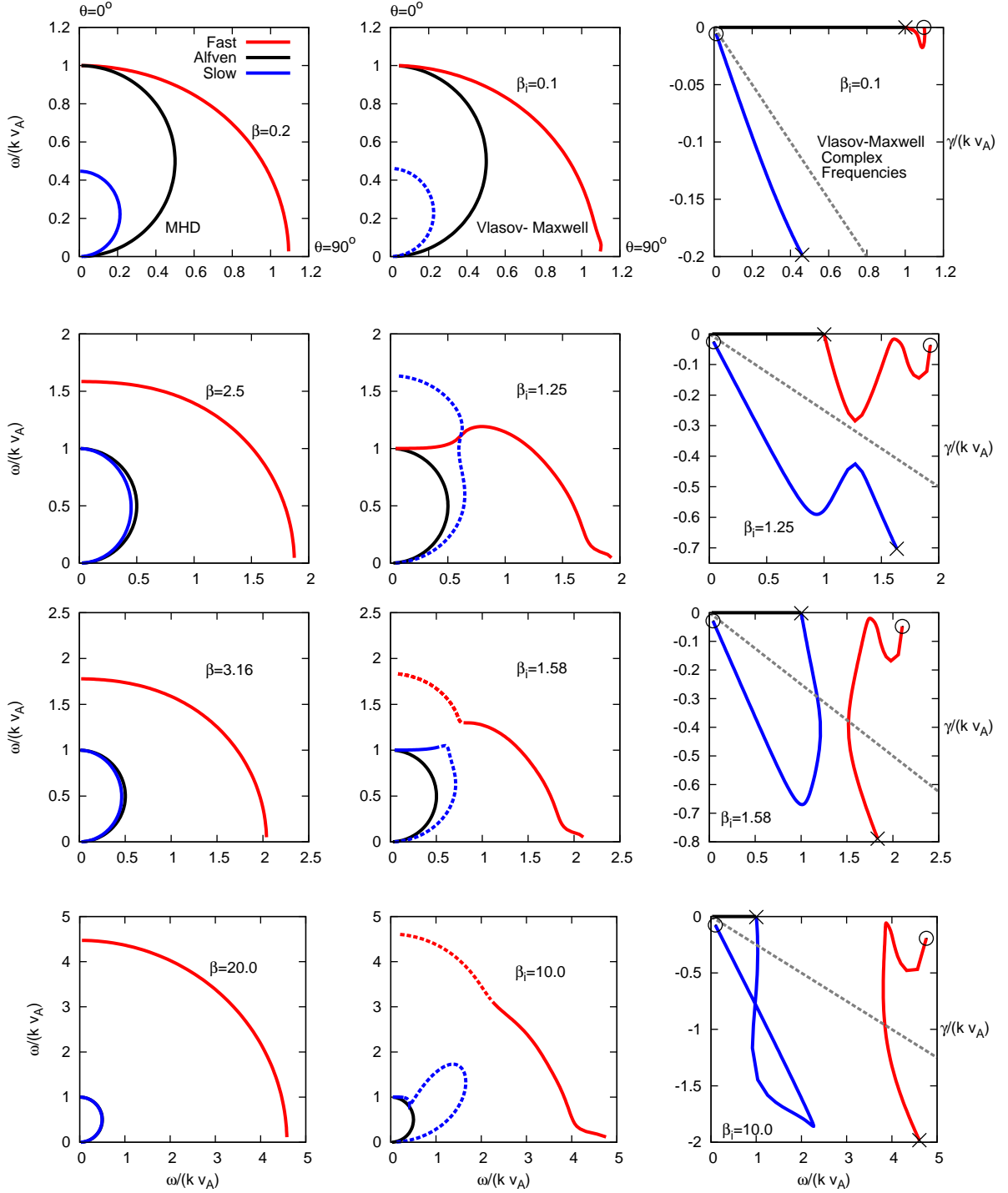} 
\caption{ Fredrichs Diagrams of normalized phase speed 
$\omega/(k v_A)$ for MHD (left column) and Vlasov-Maxwell (middle
column) modes for $k \rho_i =2.2 \times 10^{-4}$ and values of $\beta = 2\beta_i
= $ 0.2, 2.5, 3.16, and 20.  Parallel wavevectors, $\theta=0^\circ$ with
respect to the local mean magnetic field, lie along the the vertical axis,
while perpendicular wavevectors, $\theta=90^\circ$, lie along the
horizontal axis.  Dashing indicates $-\gamma / \omega > 0.25$, which
implies heavy collisionless damping of that wave mode.  On the right are
the paths of the Vlasov-Maxwell solutions in complex $\omega$ space,
running from the nearly parallel limit (crosses) to the nearly
perpendicular limit (open circles), with $\gamma / \omega = -0.25$
plotted as a dashed grey line.\label{fig:Polar} }
\end{figure}

Previous studies have compared the properties of linear Vlasov-Maxwell
wave modes with the linear modes from two-fluid theory
\citep{Krauss-Varban:1994} and Hall MHD \citep{Howes:2009b}.  Here we
restrict ourselves to establishing the connection between MHD wave
modes and Vlasov-Maxwell wave modes in the inertial range limit, $k
\rho_i \ll 1$. Due to the reduced parameter space of the
Vlasov-Maxwell system in the inertial range limit, the comparison
between the three linear MHD wave modes and their kinetic counterparts
is concisely expressed by the use of normalized Freidrichs diagrams,
polar plots of the wave phase velocity normalized to the \Alfven speed
$\omega/(k v_A)$ vs. polar angle $\theta$, as shown in
\figref{fig:Polar} for four values of $\beta_i$.

As discussed by \citet{Krauss-Varban:1994}, the identification of
corresponding wave modes is complicated by the existence of a branch
cut in the complex solution space of the Vlasov-Maxwell dispersion
relation. At inertial range scales $k \rho_i \ll 1$, this branch cut
exists only at small angles $\theta$; for wavevectors in the
perpendicular limit, the kinetic fast, Alfv\'en, and slow modes are
always easily distinguished.  Therefore, we adopt the strategy that we
label the modes according to their properties in the perpendicular
wavevector limit, and retain those labels for the linear dispersion
relation solutions as the angle is decreased from $\theta= 90^\circ$
to $\theta= 0^\circ$. Further discussion of the properties of the
kinetic fast and slow wave modes at small angles $\theta$ is presented
in Appendix~\ref{app:branch}.

Results of the MHD to Vlasov-Maxwell comparison are presented in
\figref{fig:Polar}. In the left column are the normalized Freidrichs
diagrams for the MHD results for the fast wave (red), \Alfven wave
(black), and slow wave (blue). In the center column are the
Vlasov-Maxwell results for the kinetic fast wave (red),
\Alfven wave (black), and kinetic slow wave (blue). Note that the strongly
damped Vlasov-Maxwell modes (defined by $-\gamma/\omega > 0.25$, where
$\omega+ i \gamma$ is the complex Vlasov-Maxwell eigenfrequency) are
given by dashed lines. In the right column are the paths of the
Vlasov-Maxwell solutions in complex $\omega$ space, running from the
nearly parallel limit (crosses) to the nearly perpendicular limit
(open circles).  The region below the grey dashed line is
$-\gamma/\omega > 0.25$, indicating strong collisionless damping. Rows
present different values for ion plasma beta, $\beta_i=0.1, 1.25,
1.58, 10.0$.

For most values of $\beta_i$ and $\theta$ presented in the leftmost
two columns of \figref{fig:Polar}, the correspondence between the MHD
and kinetic wave modes is clear. It is worth noting, however, several
distinctions between the fluid and kinetic behavior.  First, the
kinetic slow wave has a greater phase speed than the \Alfven wave for
$\beta_i \gtrsim 1$.  In fact, for sufficiently parallel wavevectors and
$\beta_i \sim 1$, the kinetic slow mode also has a greater phase
velocity than the fast mode.  Second, the typical magnetic-to-acoustic
mode conversion that occurs for the MHD fast and slow modes at $\beta
= 1$ is replaced by a conversion between the compressible kinetic
roots at $\beta_i \sim 1.3$.  This conversion leads to damping of the
fast waves at small angles $\theta$ in the large $\beta_i$ limit.
This general identification of the kinetic fast and slow wave modes
is used in the construction of synthetic spacecraft data, as
described in \secref{sec:ssd}.

\subsection{Density-Parallel Magnetic Field Correlation for Kinetic Fast and Slow Modes}
\label{sec:nbpar}

With a clearly defined identification of the kinetic fast and slow
modes complete, we may now calculate the density-parallel magnetic
field correlation for the kinetic fast and slow modes.  This will
enable us to verify whether the general qualitative properties of this
correlation---that fast waves are positively correlated, and slow
waves negatively correlated---remain unchanged for collisionless
conditions. The density fluctuation $\delta n$ and parallel magnetic
field fluctuation $\delta B_\parallel$ may be calculated from the
complex solution of the linear Vlasov-Maxwell dispersion relation. For
a chosen wavevector $\V{k}$, we obtain complex Fourier coefficients
$\delta n (\V{k})$ and $\delta B_\parallel(\V{k})$, and we define the
normalized correlation by Fourier mode, $C_\V{k}$, given by
\begin{equation}
C_\V{k}(\delta n , \delta B_\parallel) = \mbox{Re} \left(\frac{\delta n (\V{k})\delta B_\parallel(\V{k})^*}
{|\delta n (\V{k})||\delta B_\parallel(\V{k})|}\right).
\end{equation}
Note here that the Fourier coefficients of both density and parallel
magnetic field fluctuations satisfy the reality condition, \emph{e.g.},
$\delta n (\V{k}) =\delta n^* (-\V{k})$.

\begin{figure}[t] 
   \includegraphics[viewport=5 0 440 100,clip,scale=1.15]{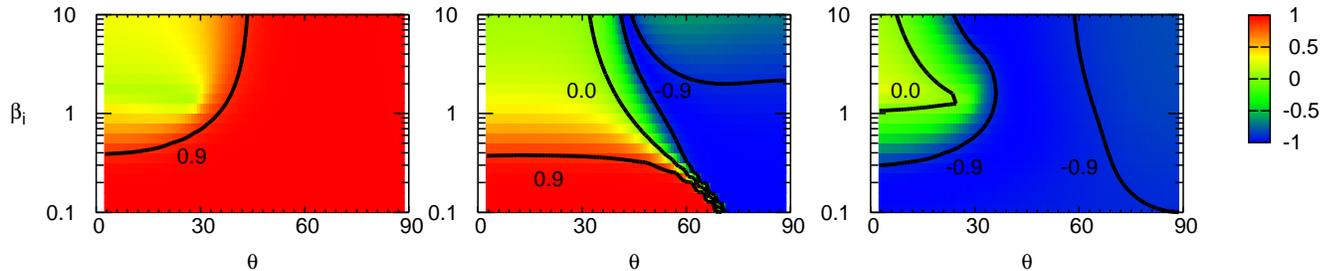}
\caption{ The normalized correlation by Fourier mode, 
$ C_\V{k}(\delta n , \delta B_\parallel)$, for the kinetic fast
(left), \Alfven (center), and slow (right) modes over the plane
$(\theta,\beta_i)$. A positive correlation is found for the kinetic
fast modes, and a generally negative correlation for the slow modes.}
\label{fig:proxy}
\end{figure}

In \figref{fig:proxy}, $ C_\V{k}(\delta n , \delta B_\parallel)$ is
plotted for the kinetic fast (left), \Alfven (center), and kinetic
slow (right) modes as a function of the parameters $(\theta,\beta_i)$.
The remaining parameters for the Vlasov-Maxwell eigenfunction
solutions are $k \rho_i =2 \times 10^{-4}$, $T_i/T_e=1$, and
$v_{t_i}/c =10^{-4}$.  The kinetic fast mode (left) always has a
positive correlation, with $ C_\V{k}(\delta n , \delta
B_\parallel)>0.9 $ over most of the $(\theta,\beta_i)$ plane.  The
kinetic slow mode (right) has a negative correlation with $
C_\V{k}(\delta n , \delta B_\parallel)< -0.9 $ over most of the
$(\theta,\beta_i)$ plane; the correlation becomes slightly positive
for $ \theta < 30^\circ$ and $\beta_i
\geq 1.3$. The \Alfven mode (center), presented for completeness, 
has a more complicated behavior, but it is worthwhile noting that the
amplitudes of $\delta n (\V{k})$ and $\delta B_\parallel(\V{k})$ are
both very small (compared to the characteristic amplitudes for either
of the compressible waves) at inertial range scales, so this
correlation for \Alfven waves is likely to be unmeasurable in the
solar wind.  In general, these results confirm that the general
qualitative properties of the density-parallel magnetic field
correlation remain unchanged for the kinetic fast,
Alfv\'en, and slow modes in a weakly collisional plasma.

\section{Constructing Synthetic Spacecraft Data}
\label{sec:ssd}

The primary methodology employed in this study of the compressible
fluctuations in the solar wind inertial range is the construction of
synthetic spacecraft data for direct comparison to actual single-point
spacecraft measurements. For the present study of the correlation
between the density and parallel magnetic field fluctuations,
this technique enables us to determine the characteristics of the correlation
in the presence of a turbulent spectrum of wave modes. This method of
analysis is based on the \emph{quasi-linear premise} that some
characteristics of magnetized plasma turbulence can be usefully
modeled as a collection of randomly phased, linear wave modes; the
justification for this premise is discussed in
\secref{sec:premise}. The result from \secref{sec:model}---that, in 
the inertial range, the dynamics of both MHD and Vlasov-Maxwell plasmas
depend only on two parameters, the ion plasma beta $\beta_i$ and the
wavevector angle $\theta$---significantly simplifies the construction
of synthetic plasma data. Upon adoption of the quasi-linear premise
and specification of the plasma $\beta_i$, one need only choose the
partitioning of power among the contributing linear wave modes (fast,
Alfv\'en, or slow) and the wavevector distribution of power (isotropic
or critically balanced) for each of those modes, as described in
\secref{sec:procedure}.

\subsection{The Quasi-linear Premise: Modeling Turbulence as a 
Spectrum of Linear Wave Modes}
\label{sec:premise}

The theoretical investigation of turbulence using synthetic data is
based on a concept that we denote the \emph{quasi-linear premise}.
The quasi-linear premise \footnote{Note that the quasi-linear premise
is not the same as, and does not require, the quasi-linear
approximation, a rigorous mathematical procedure that requires weak
nonlinear interactions such that perturbation theory can be applied.}
states that some properties of magnetized plasma turbulence can be
understood by modeling the turbulence as a collection of randomly
phased, linear waves. In this picture, the nonlinear turbulent
interactions serve to transfer energy from one linear wave mode to
another---thus, the picture is quasi-linear. The mathematical
properties of the equations that describe turbulence in a magnetized
plasma, in conjunction with a phenomonological understanding of the
properties of the turbulence, provide the motivation for this
quasi-linear approach.  This concept can be most easily explained
using the following example of turbulence in an incompressible MHD
plasma.

The Elsasser form  of the ideal incompressible
MHD equations \citep{Elsasser:1950} is given by
\begin{equation}
\frac{\partial \V{z}^\pm}{\partial t} \mp (\V{v}_A \cdot \nabla) \V{z} ^\pm  = - 
(\V{z}^\mp \cdot \nabla)\V{z}^\pm-\frac{\nabla p}{\rho_0},
\label{eq:elsasser}
\end{equation}
where the magnetic field has been decomposed into its equilibrium and
fluctuating parts $\V{B}=\V{B}_0+ \delta \V{B} $, the \Alfven velocity
due to the equilibrium magnetic field $\V{B}_0$ is given by $\V{v}_A
=\V{B}_0/\sqrt{4 \pi \rho_0}$, and $\V{z}^\pm = \delta \V{v} \pm \delta
\V{B}/\sqrt{4 \pi\rho_0}$ are the Elsasser fields describing the
velocity and magnetic field behavior of waves traveling down (up) the
mean magnetic field.  The second term on the left-hand side of
\eqref{eq:elsasser} represents the linear propagation of the
Elsasser fields along the mean magnetic field at the \Alfven speed,
while the terms on the right-hand side represent the nonlinear
interactions between upward and downward propagating waves, where the
pressure gradient term ensures incompressibility of the fluctuations.

The theory of strong incompressible MHD turbulence
\citep{Goldreich:1995,Boldyrev:2006} suggests that the turbulent 
fluctuations at small scales become anisotropic, where the nonlinear
cascade of energy generates turbulent fluctuations with smaller scales
in the perpendicular direction than in the parallel direction,
$k_\parallel \ll k_\perp$. This inherent anisotropy of magnetized
plasma turbulence has long been recognized from early studies in
laboratory plasmas \citep{Robinson:1971,Zweben:1979,Montgomery:1981}
and in early
numerical simulations \citep{Shebalin:1983}.  It has been conjectured
that strong turbulence in incompressible MHD plasmas maintains a state
of \emph{critical balance} between the linear timescale for \Alfven
waves and the nonlinear timescale of turbulent energy transfer
\citep{Higdon:1984a,Goldreich:1995}. There exists significant evidence
consistent with the predictions of critical balance from analysis of
numerical simulations \citep{Cho:2000,Maron:2001,Cho:2003} and solar wind
observations
\citep{Horbury:2008,Podesta:2009a,Wicks:2010,Luo:2010,Chen:2011c,Forman:2011}. 
In a state of strong turbulence, critical balance implies that the linear term
$(\V{v}_A \cdot\nabla) \V{z} ^\pm $ and nonlinear term $ (\V{z}^\mp
\cdot\nabla)\V{z}^\pm $ in \eqref{eq:elsasser} are of the same 
order\footnote{Note that, in the case of incompressible hydrodynamic
turbulence (Euler or Navier-Stokes), the absence of a linear term
prohibits the possibility of a quasi-linear approach.}.  It is this
property that motivates the adoption of the quasi-linear premise, as
discussed below.

In the absence of the nonlinear terms (setting the first term on the right-hand side of
\eqref{eq:elsasser} to zero), the behavior of the plasma 
is entirely determined by the linear term. If the right-hand side of the
equation is considered to be an arbitrary perturbing source term, the
linear term determines the instantaneous response of the plasma to the
imposed perturbation. In the case of weak turbulence
\citep{Sridhar:1994,Montgomery:1995,Ng:1996,Goldreich:1997,Ng:1997,Galtier:2000,Lithwick:2003}, the nonlinear terms on the right-hand of \eqref{eq:elsasser} 
are indeed a small perturbation to the linear system, representing the
nonlinear transfer of energy between the linear wave modes.
Perturbation theory may be applied to the study of the turbulent
dynamics in this limit, so the quasi-linear premise is clearly valid
for the case of weak turbulence.  

For strong turbulence, the condition of critical balance implies that
the energy in a particular linear wave mode may be transferred
nonlinearly to other modes on the timescale of its linear wave
period. But since the linear and nonlinear terms are of the same order
in critical balance, the linear term still contributes significantly
to the instantaneous response of the plasma, even in the presence of
strong nonlinearity. Therefore, the fluctuations in a strongly
turbulent magnetized plasma are expected to retain at least some of
the properties of the linear wave modes. In particular, for a
turbulent fluctuation with a given wavevector, the amplitude and phase
relationships between different components of that fluctuation are
likely to be related to linear eigenfunctions of the characteristic
plasma wave modes. In the construction of synthetic plasma turbulence
data, a spectrum of randomly phased linear wave modes can be
specified, with the amplitude of each of the linear modes adjusted to
satisfy a chosen observational constraint, such as the turbulent
magnetic energy spectrum (the second-order moment).  By adopting the
quasi-linear premise, the properties of the synthetic turbulence data
may then be compared directly to spacecraft measurements to explore
the nature of turbulent fluctuations.

The third- and higher-order moments of the turbulence, on the other
hand, are clearly \emph{not} described by this simplified quasi-linear
approach.  Such higher order statistics depend critically on the phase
relationships between different linear wave modes, and these phase
relationships are determined by the nonlinear interactions responsible
for the turbulent cascade of energy from large to small scales. For a
collection of randomly phased linear waves, such higher order
statistics of synthetic data constructed using the quasi-linear
premise will average to zero, yielding no useful information.

Although our illustration of the application of the quasi-linear
premise above specifies the case of turbulence in an incompressible
MHD plasma, the necessary general properties of the linear and
nonlinear terms, as well as the inherent anisotropy of magnetized
plasma turbulence, continue to hold for less restricted plasma
conditions, including kinetic plasmas
\citep{Howes:2006,Schekochihin:2009}.  In particular, the arguments
for the importance of the linear physics even in a strongly turbulent
plasma also hold for the linear collisionless damping rates, providing
the foundation for simple models of the turbulent energy cascade,
encompassing both the inertial and dissipation ranges
\citep{Howes:2008b,Howes:2011b}.

Besides the feasibility arguments for the validity of the quasi-linear
premise outlined above, we give here no \emph{a priori} proof for its
validity in strongly nonlinear plasma turbulence.  Nonlinear
simulations of plasma turbulence and observational studies of solar
wind turbulence provide two avenues for testing the validity of the
premise---at present, there exist arguments in the literature both for
and against its validity.  Analysis of nonlinear numerical simulations
of plasma turbulence using gyrokinetics
\citep{Howes:2008a,Howes:2011a} and both Hall MHD and Landau fluid
theory \citep{Hunana:2011}, as well as observational analysis of
multi-spacecraft data in the solar wind
\citep{Sahraoui:2010b}, support the  validity of the 
quasi-linear premise. In addition, given that the idea of critical
balance in strong MHD turbulence \citep{Goldreich:1995} is essentially
a quasi-linear concept---that the timescale of the nonlinear energy
transfer remains of order the \emph{linear} wave frequency---evidence
in support of critical balance
\citep{Cho:2000,Maron:2001,Cho:2003,Horbury:2008,Podesta:2009a,Wicks:2010,Forman:2011}
also indirectly supports the quasi-linear premise. In contrast,
studies of 3D incompressible MHD simulations \citep{Dmitruk:2009} and
2D hybrid simulations \citep{Parashar:2010} of plasma turbulence, as
well as an observational analysis of multi-spacecraft data in the
solar wind \citep{Narita:2011}, have called into question the validity
of the quasi-linear premise.  A review of this supporting and
conflicting evidence for the quasi-linear premise is presented in
\citet{Howes:2011f}, focusing in particular on questionable aspects 
of the conflicting studies that cast doubt on the validity of their
conclusion that the quasi-linear premise is inapplicable to the case
of strong plasma turbulence.

It is also important to note that the utility of the quasi-linear
premise for the study of plasma turbulence, however, may also be
judged \emph{a posteriori} by the insights gained from such an
approach. 

\subsection{Procedure for Constructing Synthetic Spacecraft Data}
\label{sec:procedure}

Upon adopting the quasi-linear premise, the construction and analysis
of synthetic spacecraft data requires three steps:
\begin{enumerate}
\item Populate a synthetic plasma volume with a spectrum of linear wave modes
with a chosen distribution of power in wavevector space.
\item Sample the synthetic plasma volume at the position of a probe 
moving with respect to the plasma to generate reduced time series
comparable to single-point spacecraft measurements.
\item Perform the requisite analysis on the synthetic time series 
to compare to spacecraft data analysis, for example, the zero-lag
cross-correlation of the density and parallel magnetic field
fluctuations.
\end{enumerate}
For the study of the density and parallel magnetic field correlation
of the solar wind compressible turbulent fluctuations, each of these steps
is detailed below.

\subsubsection{Creating the Synthetic Turbulent Plasma}

As discussed in \secref{sec:model}, for the inertial range of solar
wind turbulence, the appropriately normalized linear physics of both
the MHD (collisional fluid) and Vlasov-Maxwell (collisionless kinetic)
systems depends on a reduced set of two parameters: the ion plasma
beta $\beta_i$ and the wavevector angle $\theta$. Therefore, a
completely general synthetic turbulent plasma requires the
specification of just three properties: (1) the plasma beta, (2) the
fraction of power in each of the possible linear wave modes, and (3)
the distribution of power in wavevector space for each of these wave
modes. Both observational constraints and phenomenological models of
plasma turbulence guide our choices for these properties.

First, we discretize the three-dimensional wavevector space on a
uniform grid of $32^3$ points, where each wavevector component spans
$k_i \rho_i\in [-4.8 \times 10^{-2}, 4.8 \times 10^{-2}]$ with a
minimum grid spacing $k_0\rho_i =3 \times 10^{-3}$. Taking an equilibrium
magnetic field $\V{B}_0=B_0 \zhat$ and specifying the ion plasma beta
$\beta_i$, we may solve for the normalized linear frequencies
$\overline{\omega}(\V{k})$ and the linear eigenfunctions $\delta
\V{B}(\V{k}), \delta \V{E}(\V{k}),\delta \V{U}(\V{k}),\delta n(\V{k})$ for each 
wavevector $\V{k}$ using our chosen plasma description (MHD or
Vlasov-Maxwell). Note that we adopt the convention $\omega \ge 0$,
so that the direction of the wave group velocity (in the case of
\Alfven waves, up or down the mean magnetic field) is determined by
the wavevector.

Next, we specify the fraction of the turbulent magnetic power for each
Fourier component due to the combination of the fast, Alfv\'en, and
slow waves. After specifying the parititioning of turbulent power
among the linear wave modes, we adjust the amplitudes of the Fourier
coefficients of the linear wave modes so that the fluctuating magnetic
power $|\delta \V{B}(k)|^2$ (due to the linear superposition of all of
the contributing wave modes at a given wavenumber $k$) is consistent
with the inertial range observational constraint that the
one-dimensional magnetic energy spectrum scales as $E_B
\propto k^{-5/3}$. The amplitudes of the Fourier coefficients for the
remaining fields $ \delta\V{E}(\V{k})$, $\delta \V{U}(\V{k})$, and
$\delta n(\V{k})$ are specified by the eigenfunction solution of each
of the linear wave modes. Random phases are also applied to each wave
mode.

Finally, the distribution of energy in wavevector space for each of
the constituent wave modes must be specified to model the inherently
anisotropic nature of magnetized plasma turbulence. Imbalance between
the turbulent energy fluxes propagating up and down the magnetic field
as well as anisotropy in the angular distribution of turbulent energy
with respect to the mean field direction may be incorporated in this
final step. For this initial synthetic data investigation of the
compressible fluctuations in the inertial range, we always set the
upward and downward propagating wave energy fluxes in
balance. Numerical simulations of compressible MHD plasma turbulence
suggest that fast wave energy is distributed isotropically while both
the
\Alfven and slow wave energies obey a critically balanced distribution with 
energy concentrated mainly in modes with $k_\parallel \lesssim
k_0^{1/3} k_\perp ^{2/3}$, where $k_0$ is the isotropic driving scale
of the turbulence \citep{Cho:2003}. Therefore, we allow for two
possible wavevector distributions of energy: (a) an isotropic
distribution, such that the distribution of power is independent of
$\theta$; and, (b) a simplified critically balanced distribution, where all
modes with $k_\parallel > k_0^{1/3} k_\perp ^{2/3}$ are set to zero,
where $k_0$ is the minimum wavenumber of the simulation domain.

Note that the adjustment of the fraction of turbulent power due to
each wave mode is performed for each Fourier component, so that
changes in the wavevector distribution of energy---\emph{e.g.}, from
isotropic to critically balanced by zeroing out all modes with
$k_\parallel > k_0^{1/3} k_\perp ^{2/3}$---do not affect the
fluctuation amplitudes of the non-zero modes.

\subsubsection{Generating Synthetic Reduced Time Series }

Once the synthetic turbulent plasma has been completely specified in
the Fourier domain, the turbulent fields may be computed at any
position $\V{r}$ and time $t$ by summing over all contributing Fourier
modes of all constituent linear waves, for example,
\begin{equation}
\delta B_\parallel ( \V{r} ,t)  = \sum_m \sum_{\V{k}} \delta  B_{\parallel m}(\V{k})
e^{i[\V{k} \cdot \V{r} - \omega_m (\V{k}) t + \phi_{mj}]}.
\end{equation}
Here, the index $m$ indicates the contributions from fast, Alfv\'en, and
slow modes, and each contributing Fourier mode $\V{k}$ for each wave
mode $m$ is given a constant, random phase $ \phi_{mj}$, where $j=1,
\ldots, n_xn_yn_z$.  Note that the linear frequency $\omega_m (\V{k})$ for 
each constituent wave is a function of the wavevector $\V{k}$ as well as the 
wave type $m$.

Unfortunately, single-point satellite measurements do not provide full
spatial information about the turbulence for comparison to the
synthetic data.  To mimic spacecraft measurements made as the
super-\Alfvenic solar wind streams past the satellite with velocity
$\V{v}$, we sample the synthetic plasma volume at the position of a
probe moving with velocity $-\V{v}$ through the volume, $\V{r}(t) =
\V{r}_0 - \V{v}t$, where for simplicity we set $\V{r}_0=0$. Sampling 
along the probe trajectory at an interval $\Delta t$ generates
single-point time series at times $t_n=n\Delta t$ for each of the
turbulent fluctuating fields, where $n=1, \ldots N$ and the total
time interval is therefore $T=N\Delta t$. This reduced set of data is directly comparable to single-point
spacecraft measurements. The time series of the parallel magnetic
field fluctuation $\delta B_{\parallel n} \equiv \delta
B_{\parallel}(t_n)$, for example, is given by
\begin{equation}
\delta B_{\parallel n}  = \left. \delta B_\parallel ( \V{r} ,t)\right|_{\V{r}=-\V{v}t_n}  = \sum_m \sum_{\V{k}} \delta  B_{\parallel m}(\V{k})
e^{-i[(\V{k} \cdot \V{v} + \omega_m) t_n - \phi_{mj}]}.
\label{eq:notaylor}
\end{equation}

Note that the frequency $\omega'$ of the signal measured by the moving
probe is Doppler shifted by the probe velocity, $\omega'=\omega+ \V{k}
\cdot \V{v}$. Normalizing this Doppler-shifted frequency  by $k v_A$ to obtain 
 $\overline{\omega}'=\omega'/(kv_A)$, we find
 $\overline{\omega}'=\overline{\omega}+ \hat{\V{k}}
\cdot (\V{v}/v_A)$, where $\hat{\V{k}}$ is the unit vector in the direction 
of the wave vector. The Taylor hypothesis---that the temporal
fluctuations measured in the super-\Alfvenic solar wind flow are dominated
by spatial fluctuations swept past the probe at the solar wind
velocity---is equivalent to the limit $\hat{\V{k}}
\cdot (\V{v}/v_A) \gg \overline{\omega}
$ \citep{Taylor:1938}.  Because the ratio $v/v_A \simeq 10$ in the
super-\Alfvenic solar wind flow \citep{Tu:1995,Bruno:2005}, the Taylor
hypothesis is frequently a useful simplification for studies of the
non-dispersive linear wave modes of the inertial range.  For the
present study of the compressible fluctuations in the inertial range,
we adopt the Taylor hypothesis that $\omega' \simeq \V{k} \cdot
\V{v}$, so
\begin{equation}
\delta B_{\parallel n}  = \sum_m \sum_{\V{k}} \delta  B_{\parallel m}(\V{k})
e^{-i[\V{k} \cdot \V{v} t_n - \phi_{mj}]}.
\end{equation}
In \secref{sec:taylor}, we test the effect that violation of the
Taylor hypothesis has on the  density-parallel magnetic field
correlation $C(\delta n ,\delta B_{\parallel})$.

\begin{figure}[top] 
   \centering \includegraphics[width=3.5in]{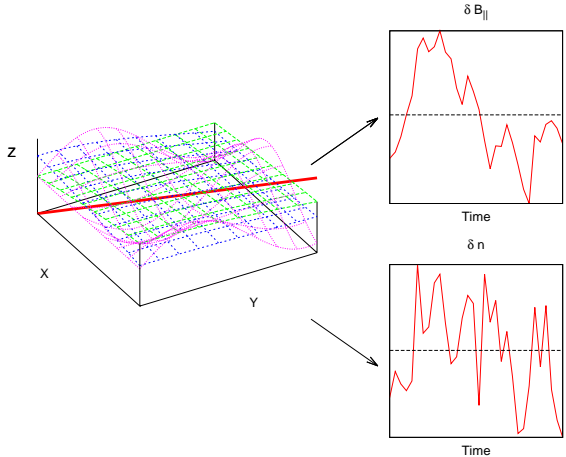}
   \caption{Probe trajectory (red) through a synthetic turbulent
   plasma volume generated using a spectrum of linear eigenfunctions
   (blue, green, and purple surface plots on the left).  Measurements
   of the plasma fluctuations along the probe trajectory yield
   synthetic single-point time series of the parallel magnetic field
   fluctuation $\delta B_{\parallel}$ (upper right) and the density
   fluctuation $\delta n$ (lower right). }  \label{fig:SatImage}
\end{figure}

This procedure of sampling the synthetic plasma volume is depicted
schematically in \figref{fig:SatImage}.  Here the blue, green, and
purple surface plots represent the spatial variation for a few of the
contributing Fourier modes to the turbulent fields. The synthetic
plasma volume is sampled at uniform time intervals along the probe
trajectory (red line), generating single-point time series of the
parallel magnetic field fluctuation $\delta B_{\parallel n}$ (upper
right) and the density fluctuation $\delta n_n$ (lower right). These
synthetic time series may then be analyzed using the same procedures as
the actual spacecraft measurements.

\subsubsection{Analysis of Synthetic Spacecraft Data}
As discussed in \secref{sec:nbpar}, the correlation between the
density fluctuation $\delta n$ and the parallel magnetic field
fluctuation $\delta B_{\parallel}$ distinguishes fast from slow
compressible modes in either the fluid MHD or kinetic Vlasov-Maxwell
systems.  The zero-lag cross-correlation of the 
two time series  $\delta
n_ n$ and $\delta B_{\parallel n}$ is given by 
\begin{equation}
C( \delta n , \delta B_{\parallel})
=\frac{\sum_n \left(\delta n_n -\overline{\delta n}\right)
	\left(\delta B_{\parallel n} -\overline{ \delta B_\parallel}\right)}{
	\sqrt{\sum_n \left(\delta n_n -\overline{\delta n}\right)^2 }
	\sqrt{\sum_n \left(\delta B_{\parallel n} -\overline{ \delta B_\parallel}\right)^2 }},
\label{eqn:corr}
\end{equation}
where $\overline{\delta n}$ and $\overline{ \delta B_\parallel}$ are
the averages over the time interval $T$. Comparison of the theoretical
predictions of $C( \delta n , \delta B_{\parallel})$ to the
correlation from spacecraft measurements, assuming the validity of the
quasi-linear premise, provides a means of constraining the nature of
the compressible fluctuations in the solar wind.

\section{Theoretical Predictions of $C( \delta n , \delta B_{\parallel})$ 
Using Synthetic Spacecraft Data}
\label{sec:ssd_predict}

Our theoretical investigation of the compressible fluctuations in the
inertial range of solar wind turbulence uses the method outlined in
\secref{sec:ssd} to  generate synthetic spacecraft data to provide
predictions of the density-parallel magnetic field correlation $C(
\delta n , \delta B_{\parallel})$ as a function of the following
turbulent plasma properties: the plasma $\beta_i$, the fraction of
fast-to-total compressible wave power $\mathcal{F}$, and the angular
distribution of the turbulent power in wavevector space. These
theoretical predictions may then be compared directly to the $C(
\delta n , \delta B_{\parallel})$ computed from satellite measurements
to constrain the nature of the compressible fluctuations.

Below we present a comparison of four methods to predict and interpret
the behavior of $C( \delta n,\delta B_{\parallel})$ as a function of
$\beta_i$ and $\mathcal{F}$: (1) an analytical estimate using MHD
eigenfunctions; (2) synthetic spacecraft data using MHD
eigenfunctions, including contributions from fast, Alfv\'en, and slow
waves; (3) synthetic spacecraft data using Vlasov-Maxwell
eigenfunctions using only the compressible kinetic fast and slow
waves; and (4) synthetic spacecraft data using Vlasov-Maxwell
eigenfunctions, including contributions from fast, Alfv\'en, and slow
waves. For each of these cases, we explore three combinations of the
angular distribution of the constituent wave mode power: (a) all
isotropically distributed wave power distributions, (b) all critically
balanced wave power distributions, and (c) isotropic fast wave and
critically balanced \Alfven and slow wave power distributions.  Based
on intuition gained from numerical simulations of compressible MHD
turbulence \citep{Cho:2003}, we expect that the third case of
isotropic fast waves and critically balanced \Alfven and slow waves is
the most realistic choice; the alternative selections are included to
demonstrate the sensitivity of the method to different power
distributions in wavevector space.

\begin{center}
\begin{figure}[p] 
\includegraphics[scale=1.1]{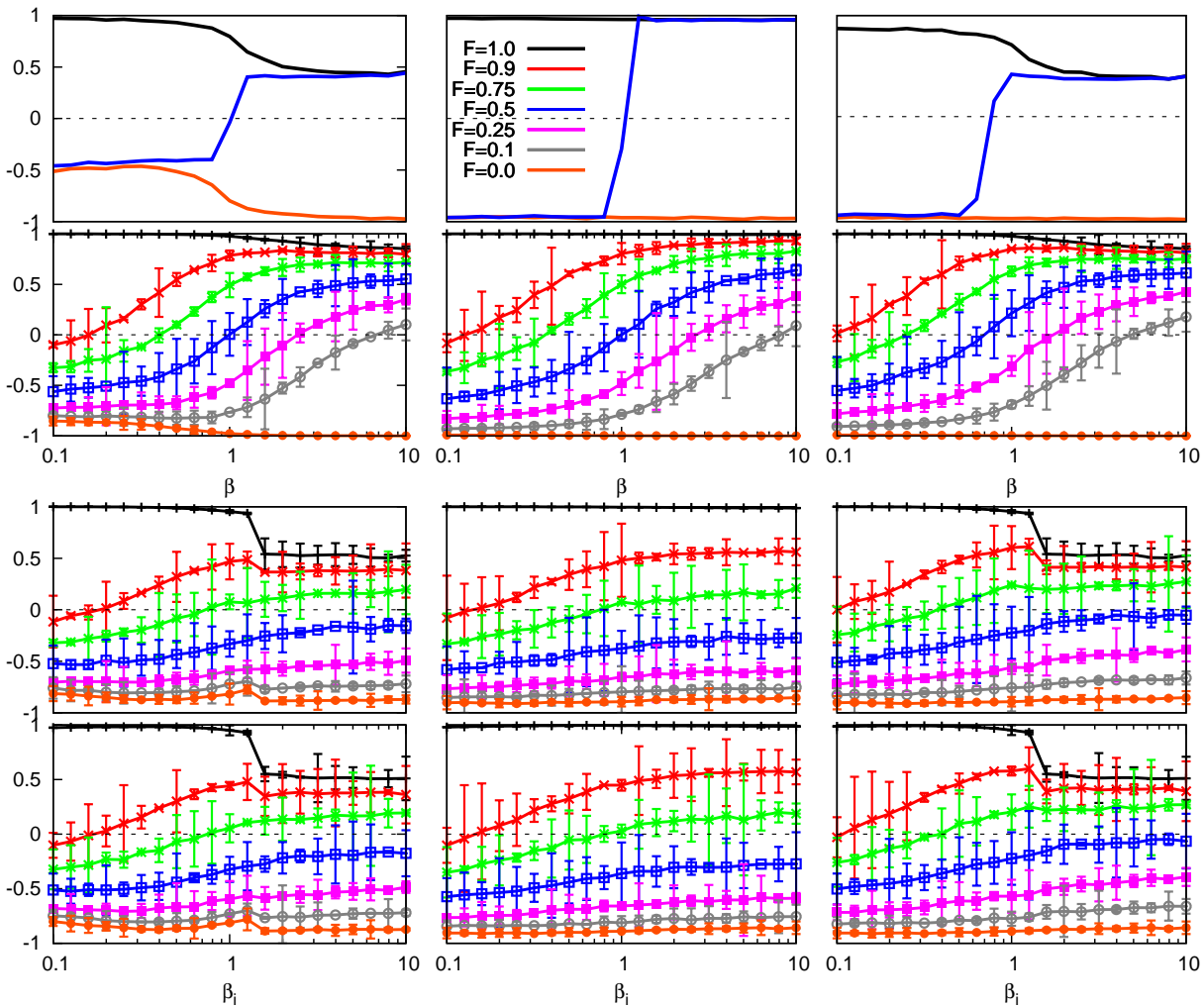}
  \caption{ Theoretical predictions of $C( \delta n , \delta
  B_{\parallel})$ using synthetic spacecraft data as a function of ion
  plasma beta $\beta_i$ and fast-to-total compressible wave power
  $\mathcal{F}$. By row are analytical predictions based on MHD  eigenfunctions
  (first), synthetic spacecraft data using MHD eigenfunctions
  (second), synthetic spacecraft data using Vlasov-Maxwell
  eigenfunctions with only fast and slow waves (third), and synthetic
  spacecraft data using Vlasov-Maxwell eigenfunctions with 90\%
  \Alfven wave power, and the remaining wave power split between fast
  and slow waves (fourth). By column are presented different angular
  distributions of wave power: all isotropic (first), all critically
  balanced (second), isotropic fast and critically balanced \Alfven
  and slow (third).  Results are presented for varying fractions of
  fast-to-total compressible wave power $\mathcal{F}$: 1.0 (black),
  0.9 (red), 0.75 (green), 0.5 (blue), 0.25 (magenta), 0.1 (grey), 0.0
  (orange).  The error bars are the standard deviation from each
  ensemble of runs.} \label{fig:ssd_nbpar}
\end{figure}
\end{center}

All of the synthetic spacecraft data results of $C(\delta n, \delta
B_{\parallel})$ in this section are relatively insensitive to the
angle of the probe trajectory with respect to the equilibrium magnetic
field, as long as the trajectory is not exactly parallel or
perpendicular to the field; all results here used an angle of
$45^\circ$ with respect to both the equilibrium magnetic field and the
$x$-axis. As a test of the validity of the Taylor hypothesis, we
created sets of syntehtic data from a time evolving plasma in which
the Taylor hypothesis is not assumed.  The time evolution of each
Fourier component for each wave mode was prescribed by its linear
frequency and the rate of time evolution is  parametrized by the ratio of
solar wind to \Alfven velocities $v/v_A$.  The correlation $C(\delta
n, \delta B_{\parallel})$ from the time evolved and stationary cases
were indistinguishable as long as the motion is sufficiently super-
\Alfvenic ($v/v_A\gtrsim 10$), which is typically satisfied in the
solar wind \citep{Tu:1995,Bruno:2005}.  Therefore, we assume the
Taylor hypothesis is valid and present results derived from stationary
synthetic data sets.  For each choice of $\beta_i$ and $\mathcal{F}$,
the values of $C(\delta n, \delta B_{\parallel})$ plotted in
\figref{fig:ssd_nbpar} are the mean of 256 ensembles; the error bars
are the standard deviation from this statistical averaging procedure.
The ensembles are used to average over the random phases of the 
fluctuations.

\subsection{Analytical Estimate of $C( \delta n , \delta B_{\parallel})$} 
To build our intuition about the behavior of $C( \delta n , \delta
B_{\parallel})$ for turbulence modeled as a spectrum of randomly
phased linear wave modes, we use the normalized compressible MHD
eigenfunctions presented in
\appref{app:MHDeig} to construct an analytical estimator $\Xi(\beta)$ for 
$C( \delta n , \delta B_{\parallel})$. For a plasma with an
equilibrium magnetic field $\V{B}_0 =B_0 \zhat$, the density and
parallel magnetic field fluctuation for a Fourier mode with wavevector
$\V{k} = k \sin \theta \xhat + k \cos \theta \zhat$ are generally
given by the normalized Fourier coefficients $\delta
\hat{n}( \beta,\theta)$ and $\delta \hat{B}_\parallel( \beta,\theta)$. 
We can estimate the correlation by integrating the correlations of the
Fourier coefficients over the specified angular distributions of power for each
wave mode,
\begin{equation}
\Xi(\beta)=\frac{ \int \sum_m \left[\delta
\hat{n}_m( \beta,\theta)\right] \sum_m \left[\delta \hat{B}_{\parallel m}( \beta,\theta)\right] d \theta }
{\sqrt{(\int \sum_m \left[\delta
\hat{n}_m( \beta,\theta)\right] d \theta)^2}
\sqrt{(\int \sum_m \left[\delta \hat{B}_{\parallel m}( \beta,\theta)\right]  d \theta)^2}},
\label{eqn:Xi}
\end{equation}
where the sum over $m$ includes the contribution from each of the
constituent wave modes.  The expected dependence on $\beta$ is
explicit in the expression, while the choice of mode fraction
$\mathcal{F}$ is implicit in the choice of the fraction of power in
each of the contributing wave modes.  Since the \Alfven wave mode in
the MHD limit has Fourier coefficients $\delta\hat{n} = \delta
\hat{B}_\parallel=0$, it is unnecessary to include a contribution from
 \Alfven waves to the estimated correlation $\Xi(\beta)$.
In the top row of \figref{fig:ssd_nbpar}, $\Xi(\beta)$  is plotted for
the three angular power distribution cases: all isotropic (left), all
critically balanced (center), and isotropic fast and critically
balanced slow (right). For each case, we compute the estimator
$\Xi(\beta)$ for three fractions of fast-to-total compressible wave
power $\mathcal{F}=0$ (orange), $\mathcal{F}=0.5$ (blue), and
$\mathcal{F}=1$ (black). 

Two qualitative features are immediately apparent from the top row of
\figref{fig:ssd_nbpar}.  First, the behavior for the mixture of fast
and slow modes is dominated by the slow mode in the $\beta < 1$ region
and by the fast mode in the $\beta > 1$ region.  This result can be
understood by comparing the magnitude of the fast and slow mode density
fluctuations in these regions.  In the small $\beta$ region, the slow
mode density fluctuations are one to two orders of magnitude larger
than the fast mode fluctuations; the opposite case holds for the high
$\beta$ region, where the density fluctuations for the fast mode are
much larger than for the slow mode. Second, for the case of both
isotropic wave power distributions (left), we see a dependence on
$\beta$ both for pure fast modes at $\beta \gtrsim 1$ and for pure
slow modes at $\beta \lesssim 1$.  Noting that these pure modes have
practically no $\beta$ dependence for the critical balance case
(center), we can surmise that the parallel modes are the cause of the
deviations seen in the isotropic case. The case of the combination of
isotropic fast and critically balanced slow modes (right) appears to
confirm this finding because the $\beta \lesssim 1$ (slow-wave
dominated) region looks like the purely critically balanced case,
while the $\beta \gtrsim 1$ (fast-wave dominated) region appears very
much like the purely isotropic case.  These qualitative
characteristics of the analytical estimator $\Xi(\beta)$ for 
$C(\delta n, \delta B_{\parallel})$ provide a foundation upon which to
interpret the synthetic spacecraft data predictions.

\subsection{Synthetic Spacecraft Data Prediction of $C( \delta n , \delta B_{\parallel})$
using MHD Fast, Alfv\'en, and Slow Eigenmodes} 

The compressible MHD eigenfunctions presented in
\appref{app:MHDeig} are used in the procedure outlined in \secref{sec:ssd}
to generate synthetic spacecraft data to theoretically predict the
behavior of $C(\delta n, \delta B_{\parallel})$ as a function of the
synthetic turbulent plasma properties. The synthetic plasma volume is
sampled at $N=32$ uniformly spaced points along a trajectory of length
$L = vT = \frac{2 \pi}{3} \times 10^3 \rho_i$. Although the MHD
\Alfven wave has $\delta\hat{n} = \delta \hat{B}_\parallel=0$, the \Alfven wave
contribution to the synthetic turbulent plasma is included for
completeness, with 90\% of the turbulent magnetic power given by
incompressible \Alfven waves. The remaining 10\% of the magnetic power
is split between the MHD fast and slow waves according to the
specified fraction of fast-to-total compressible wave power
$\mathcal{F}$. Tests have shown that the results for $C(\delta n ,
\delta B_{\parallel})$ using MHD eigenfunctions are unaffected by the
presence or absence of the \Alfven wave contribution, as
expected. The resulting zero-lag cross correlation of the density and
parallel magnetic field $C(\delta n ,
\delta B_{\parallel})$ from the synthetic MHD plasma is presented in
the second row of \figref{fig:ssd_nbpar}.

We see that $C(\delta n , \delta B_{\parallel})$ from the MHD
synthetic data displays the same qualitative behavior as the
analytical estimate $\Xi (\beta)$.  The dominance in the low (high)
$\beta$ regions by the slow (fast) mode behavior is evident.  A plasma
that has 90\% of its compressive energy in the fast mode has a
slightly negative correlation at $\beta=0.1$, while a 90\% slow mode
plasma has a small but positive correlation for $\beta=10$.  Both of
these values are drastically different from $C(\delta n , \delta B_{\parallel})$
at $\beta =1.0$.  This
dominant behavior holds true for all three choices of power
distributions.  In comparing the results from these distributions, we
see the marked, and expected, lack of $\beta$ dependence for the pure
fast and slow modes in the critical balance cases.

\subsection{Synthetic Spacecraft Data Prediction of $C( \delta n , \delta B_{\parallel})$
using only Kinetic Fast and Slow Vlasov-Maxwell Eigenmodes}

Unlike the MHD case, the \Alfven wave in the Vlasov-Maxwell kinetic
theory has a small but non-zero fluctuating density and parallel
magnetic field component in the inertial range limit, $k \rho_i \ll
1$.  To illuminate the contribution of each of the linear kinetic wave
modes to $C(\delta n , \delta B_{\parallel})$, in this section we
generate a synthetic plasma consisting of a spectrum of only kinetic
fast and slow wave fluctuations; in the next section, the (mostly
incompressible) \Alfvenic contribution will be included for
completeness. 

The sampling of the synthetic plasma volume is the same as for the MHD
case, using $N=32$ and $L =\frac{2 \pi}{3} \times 10^3
\rho_i$. Numerical computation of the linear Vlasov-Maxwell
eigenfunctions uses the parameters $T_i/T_e=1$,
$\frac{v_{t_i}}{c}=10^{-4}$, and $m_i/m_e=1836$, with Bessel function
sums evaluated to 100~terms to ensure accurate results
\citep{Quataert:1998,Howes:2006}.  We have assumed isotropic
Maxwellian distribution functions for protons and electrons for all
Vlasov-Maxwell synthetic spacecraft data results presented in this
paper; further exploration of anisotropic temperature distributions
may prove fruitful but is beyond the scope of the present work.

The results for $C(\delta n , \delta B_{\parallel})$ from the
Vlasov-Maxwell synthetic plasma with only kinetic fast and slow waves
are presented in the third row of \figref{fig:ssd_nbpar}. Comparing to
the MHD results (second row), we note an apparent similarity in the
qualitative behavior but also some noticeable quantitative
differences.  First, the slow mode dominance in the $\beta_i \lesssim
1$ region is still observed for the kinetic plasma, but the $\beta_i
\gtrsim 1$ region is not dominated by the fast mode as in the MHD case.
In fact, the mixed modes have very little dependence on $\beta_i$ for
$\beta_i \gtrsim 1.3$. Second, for pure fast and slow modes, the
correlation for the isotropic case (left) does have a $\beta_i$
dependence in the high and low $\beta_i$ regions respectively, as is
seen in the MHD case. However, for the pure slow mode, $C(\delta n ,
\delta B_{\parallel})$ is never perfectly anti-correlated,
reaching  a minimum value of $\simeq -0.9$. Third, as with the MHD
plasma, the pure fast modes are perfectly correlated in the low
$\beta_i$ region and become drastically less correlated for $\beta_i$
values above the mode conversion (see \appref{app:branch}) at $\beta_i
\simeq 1.3$.

\subsection{Synthetic Spacecraft Data Prediction of $C( \delta n , \delta B_{\parallel})$
using Kinetic Fast, Alfv\'en, and Slow Vlasov-Maxwell Eigenmodes}
\label{sec:fulldist}

In this section, we incorporate an \Alfven wave contribution composing
90\% of the turbulent magnetic power into the turbulent synthetic
plasma, with the remaining 10\% split between the kinetic fast and
slow modes.  Note that the total compressible wave power used to
calculate $\mathcal{F}$ includes \emph{only} the kinetic fast and slow
wave contributions and does not include the small contribution to the
compressible energy from the \Alfven wave. Since the density and
parallel magnetic field fluctuations of the \Alfven wave both have
small amplitudes, it is expected that the addition of the \Alfvenic
component will not yield significant quantitative changes in $C(
\delta n , \delta B_{\parallel})$.

Sampling of the synthetic plasma volume and computation of the linear
Vlasov-Maxwell eigenfunctions is the same as in the previous section.
The resulting $C(\delta n , \delta B_{\parallel})$ for the
Vlasov-Maxwell synthetic plasma with kinetic fast, Alfv\'en, and slow
waves is presented in the fourth row of \figref{fig:ssd_nbpar}. In
comparison to the third row of \figref{fig:ssd_nbpar}, in which the
90\% of \Alfvenic fluctuation energy is not included, it is clear that
the inclusion of the dominant \Alfvenic component of the turbulence
does not lead to significant quantitative changes in $C( \delta n ,
\delta B_{\parallel})$.

\subsection{Sensitivity of  $C( \delta n , \delta B_{\parallel})$
to Violation of the Taylor Hypothesis}
\label{sec:taylor}
To test the effect of violating the Taylor hypothesis
\citep{Taylor:1938} on our results for $C( \delta n ,
\delta B_{\parallel})$, we construct a new time series from a 
temporally evolving plasma where the Taylor hypothesis is \emph{not}
assumed. By abandoning the Taylor hypothesis, we are forced to use
\eqref{eq:notaylor} to compute the time series for each of the fields 
computed using our synthetic data method. In this case, each
contributing wave mode varies with the appropriate linear frequency
$\omega_m (\V{k})$ for the particular wavevector $\V{k}$ and wave type
$m$. For a particular wave mode with plasma-frame frequency $\omega$
and wavevector $\V{k}$ sampled by a moving probe, the normalized,
Doppler-shifted wave frequency, given by
$\overline{\omega}'=\overline{\omega}+( v/v_A) \cos \theta_{kv}$,
depends on three dimensionless quantities: the normalized linear wave
frequency $\overline{\omega}= \omega/(k v_A)$, the ratio of the probe
velocity to the \Alfven velocity $v/v_A$, and the angle $\theta_{kv}$
between the wavevector of a particular mode $\V{k}$ and the probe
velocity $\V{v}$, such that $ \V{k}\cdot \V{v} = k v \cos
\theta_{kv}$.

\begin{figure}[t]
\begin{center}\resizebox{4in}{!}{\includegraphics{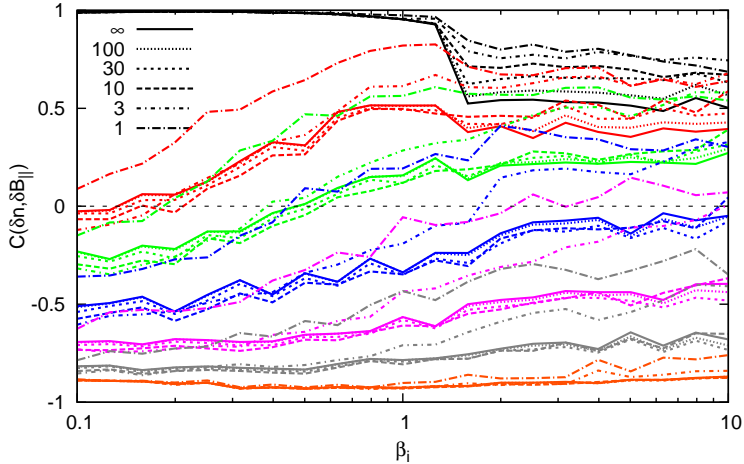}}
\end{center}
\caption{ \label{fig:notaylor} 
Theoretical prediction of $C( \delta n , \delta B_{\parallel})$ using
synthetic spacecraft data as a function of ion plasma beta $\beta_i$
and fast-to-total compressible wave power $\mathcal{F}$ when the
Taylor hypothesis is not assumed.  Synthetic spacecraft data is
produced using Vlasov-Maxwell eigenfunctions with 90\% \Alfven wave
power, and the remaining wave power split between fast and slow waves.
Variation of the ratio of the probe to \Alfven velocity tests the
sensitivity of the results to the Taylor hypothesis, where we have
tested values $v/v_A=\infty, 100,30,10,3,1$, where $v/v_A= \infty$
corresponds to the Taylor hypothesis.}
\end{figure}

For a synthetic model given by particular spectrum of linear wave
modes, each with a given distribution of power in wavevector space, once the
direction of the probe trajectory is specified, the only remaining
variable is the ratio of the probe velocity to the \Alfven velocity
$v/v_A$.  Therefore, for the same synthetic plasma model as specified
in \secref{sec:fulldist}, we may simply vary the value of $v/v_A$ to
observe the effect that violating the Taylor hypothesis has on $C(
\delta n , \delta B_{\parallel})$. The Taylor hypothesis
corresponds to $v/v_A= \infty$, whereas solar wind flows in the
near-Earth environment typically have $v/v_A \simeq 10$
\citep{Tu:1995,Bruno:2005}. In \figref{fig:notaylor}, we  plot   
$C( \delta n , \delta B_{\parallel})$ for values $v/v_A=\infty,
100,30,10,3,1$ for a probe velocity travelling at 45$^\circ$ with
respect to the mean field direction.  For values of $v/v_A \gtrsim
10$, the quantitative effect on the correlation is negligible for most
of the $(\beta_i,\mathcal{F})$ parameter space, with the exception of
slight quantitative changes for the fast wave dominated cases
$(\mathcal{F}\ge 0.9)$ at $\beta_i > 1$. Even for values as low as
$v/v_A =1 $, the qualitative appearance of the correlation $C(\delta n
, \delta B_{\parallel})$ vs. the ion plasma beta $\beta_i$ and the
fast-to-total compressible wave power $\mathcal{F}$ is essentially
unchanged.  Therefore, we conclude that the violation of the Taylor
hypothesis has little effect on the results of this investigation.

\section{Comparison to Observational Results and Discussion}
\label{sec:discuss}

Observational constraints suggest that turbulent magnetic power in
inertial range turbulence in the solar wind consists of approximately
90\%
\Alfven waves, and the remaining 10\% of the power in some mixture 
of the compressible kinetic fast and slow waves
\citep{Tu:1995,Bruno:2005}. Numerical simulations of compressible MHD
turbulence suggest that the distribution of turbulent power in wavevector space 
is isotropic for the fast waves and critically balanced for the 
\Alfven and slow waves \citep{Cho:2003}. Therefore, we believe the most 
realistic model of solar wind turbulence using synthetic spacecraft
data is given by the lower right-hand plot in \figref{fig:ssd_nbpar}.

\begin{figure}[t]
\begin{center}\resizebox{7in}{!}{\includegraphics{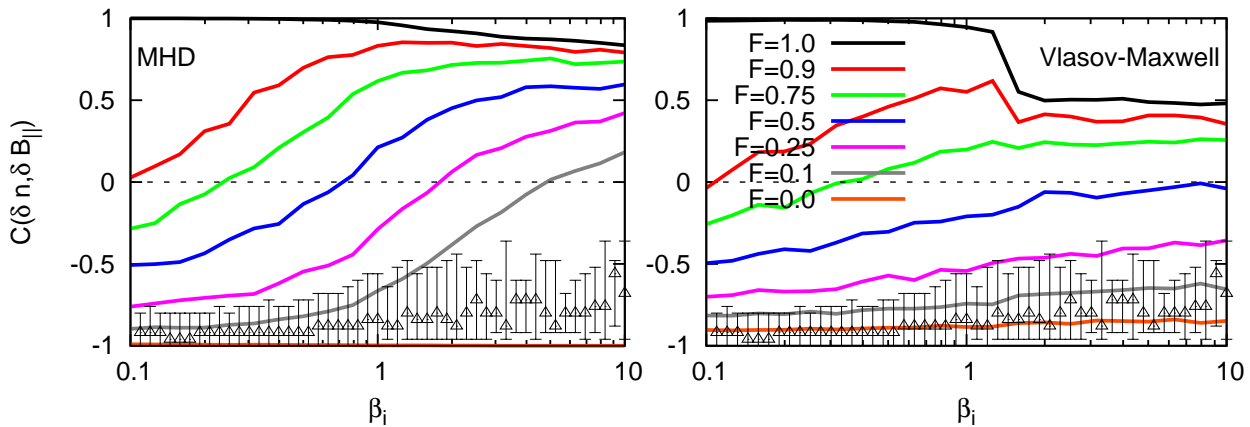}}
\end{center}
\caption{ \label{fig:money} Comparison of measured values of the 
$C(\delta n, \delta B_\parallel)$ cross-correlation (black dots with
FWHM error bars) to the synthetic data predictions for MHD (left) and kinetic
(right) theory given the ratio of
fast wave to total compressible energy $\mathcal{F}$.  Best agreement is
with the kinetic theory for $\mathcal{F}=0.00$, indicating that the compressible component of solar
wind turbulence is almost entirely in the kinetic slow mode.  The compressive behavior is not
well described by MHD for any $\mathcal{F}$.}
\end{figure}

The analysis of the density-parallel magnetic field correlation using 10 years
of \emph{Wind} spacecraft data is discussed in detail in a companion work
\citep{Howes:2012a}, so we give here only a few brief details.
The density-parallel magnetic field correlation $C(\delta n , \delta
B_{\parallel})$ is computed using measurements from the Magnetic Field
Investigation (MFI)
\citep{Lepping:1995p6447} and the Three Dimensional Plasma (3DP)
experiment \citep{Lin:1995p934} on the \emph{Wind} spacecraft in the
unperturbed solar wind at 1 AU during the years 1994-2004. Using
300-s intervals of ambient solar wind data (corresponding to inertial
range scales of approximately $k \rho_i \in (5 \times 10^{-3}, 5
\times 10^{-2})$), the proton density and magnetic field measurements
at 3~s cadence are decimated by a factor of 10 (to 30~s
cadence). Magnetic field measurements are rotated to a field-aligned
coordinate system, defined by the local mean field direction computed
using 100-s windows, to compute $\delta B_{\parallel}$, and proton
density data $\delta n = n - n_0$ is detrended over the same time
intervals. The zero-lag cross-correlation $C(\delta n ,
\delta B_{\parallel})$ is computed  for  119,512 data intervals. A joint histogram of
$C(\delta n ,\delta B_{\parallel})$ normalized in each $\beta_i$ bin
is generated, and the peak histogram values and FWHM error bars are
plotted on top of the theoretical synthetic spacecraft data plots 
of $C(\delta n ,\delta B_{\parallel})$ from MHD (left) and kinetic (right) theory 
in \figref{fig:money}.

For this direct comparison to the data, the theoretical results from
the synthetic spacecraft data are computed in the same manner as the
spacecraft measurements. In particular, the synthetic plasma volume is
sampled at only $N=10$ uniformly spaced points along a trajectory of
length $L = vT = \frac{10 \pi}{24} \times 10^2 \rho_i$, which
corresponds to a range of scales $k \rho_i \in (4.8 \times 10^{-3},
4.8\times 10^{-2})$.  The smaller number of
timesteps and shorter total sampled length leads to a larger standard
deviation using 256 ensembles than the Vlasov-Maxwell results in
\secref{sec:ssd_predict} and to slight quantitative changes in the  
$C(\delta n ,\delta B_{\parallel})$ curves.

The agreement between the $C(\delta n ,\delta B_{\parallel})$ from 
\emph{Wind} spacecraft data and synthetic data curve for $\mathcal{F}=0$  
in \figref{fig:money} is striking, indicating that the observed
correlation is consistent with a statistically negligible kinetic fast
wave energy contribution for the large sample used in this study
\citep{Howes:2012a}. We do note that a very small  fraction  of the intervals have 
$C(\delta n ,\delta B_{\parallel})>0$, possibly indicating a minority
population of fast waves.  As discussed in our companion paper
\citep{Howes:2012a}, this result has important consequences for the
turbulent cascade of energy from large to small scales: since only the
fast wave turbulent cascade is expected to nonlinearly transfer energy
to whistler waves at $k \rho_i \gtrsim 1$, and the frequency mismatch
between the fast and either the \Alfven or slow waves should prevent
non-linear coupling, our analysis suggests that there is little or no
transfer of large scale turbulent energy through the inertial range
down to whistler waves at small scales.

Comparison of the observationally measured $C(\delta n ,\delta
B_{\parallel})$ in \figref{fig:money} with both the MHD and kinetic results
leads to another important conclusion of this study: the nature of the compressible
fluctuations in the solar wind inertial range is not well modeled by
MHD theory. This result is not surprising given that the solar wind is
weakly collisional at inertial range scales, so a kinetic description
is necessary to model accurately the compressible fluctuations.

Our study also enables us to constrain the wavevector distribution of
slow wave energy by comparing the observational results with the three
plots in the bottom row of \figref{fig:ssd_nbpar}. The leftmost plot
has an isotropic distribution of slow wave power and shows two
features in the $\mathcal{F}=0$ curve not seen in the observed data:
(1) a bump at $\beta_i \sim 1$ and (2) a slight increase in the
correlation value at $\beta_i \lesssim 0.3$. These quantitative
changes in the $\mathcal{F}=0$ curve appear the same with or without
the \Alfven waves at $\beta_i \lesssim 1$ (compare the left plots in
the third and fourth row of \figref{fig:ssd_nbpar}), so this effect is
not due to the isotropic distribution of \Alfven waves. Therefore,
the slow waves appear to be anisotropically distributed, and may be well described by
critical balance as suggested by theories of slow wave passive advection 
\citep{Maron:2001,Schekochihin:2009}. However, the uncertainty of the Wind measurements
cannot definitively rule out the possibility of an isotropic
distribution for the slow mode.

One of the key points of our companion work \citep{Howes:2012a} is the
interpretation that the compressible fluctuations in the solar wind
consist of a critically balanced distribution of kinetic slow wave
fluctuations. Previous analyses have generally dismissed the
possibility of kinetic slow waves because, in an isotropic Maxwellian
plasma with warm ions, the collisionless damping via free-streaming
along the magnetic field is strong \citep{Barnes:1966}.  However, the
damping rate of the slow waves for the nearly perpendicular
wavevectors of a critically balanced distribution is proportional to
the parallel component of the wavevector, $\gamma \propto
k_\parallel$. This feature can be seen in the right column of
\figref{fig:Polar}, where the slow wave damping rate (blue) is zero at
perpendicular wavevectors $\theta=90^\circ$ (open circles) and
increases as the wavevector angle decreases. For exactly perpendicular
wavevectors, the damping rate drops to zero---this perpendicular limit
of the slow wave corresponds to an undamped, non-propagating
pressure-balanced structure (PBS). It has been derived theoretically by
\citet{Schekochihin:2009} and  demonstrated numerically by \citet{Maron:2001} that 
the \Alfven wave dynamics advects and cascades the slow waves, so the energy
cascade rate of the slow waves is related not to the slow wave
frequency, but to the \Alfven wave frequency. Therefore, although the
slow wave fluctuations at the high $k_\parallel$ boundary of critical
balanced distribution (along $k_\parallel \sim k_0^{1/3}
k_\perp^{2/3}$) may suffer strong collisionless damping, the more
nearly perpendicular slow waves may be cascaded to smaller scales on
the timescale of the \Alfvenic turbulence, while the collisionless
damping of these modes remains weak. This could lead to a slow wave
energy distribution that is \emph{more} anisotropic than critical
balance, possibly with $k_\parallel \ll k_0^{1/3} k_\perp^{2/3}$.

\begin{figure}[t]
\begin{center}\resizebox{4in}{!}{\includegraphics{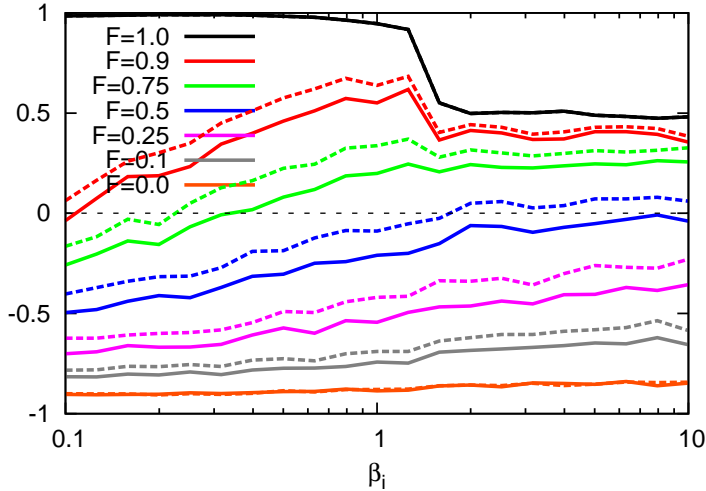}}
\end{center}
\caption{ \label{fig:ICB-ECB} Comparison of synthetic data predictions 
for $C(\delta n, \delta B_\parallel)$ with isotropic fast mode power,
critically balanced \Alfven mode power, and either critically balanced
(solid) or more anisotropic than critical balance (dashed) slow mode
power.  For the more anisotropic than critical balance distribution,
all modes with $k_\parallel > k_0 ^{2/3} k_\perp ^{1/3} $ are set to
zero, where $k_0$ is the minimum wavenumber of the wavenumber domain.
The difference between the lines is attributable to our choice for
specifying the fraction of wave power individually for each Fourier
component: the more anisotropic distribution contains less total slow wave power.
 The lines for the
$\mathcal{F} =0$ are nearly identical for both power distributions,
signifying that $C(\delta n,
\delta B_\parallel)$ can not be used to differentiate between critical
balance and more anisotropic distributions.}
\end{figure}

Using synthetic spacecraft data, we can test whether this picture of a
distribution of slow wave power that is more anisotropic than critical
balance is consistent with the observational findings in
\figref{fig:money}. To do this, we repeat the analysis in the lower right plot 
of \figref{fig:ssd_nbpar} but with a more anisotropic distribution of
slow waves given by $k_\parallel \le k_0^{2/3} k_\perp^{1/3}$. The
results of this test, shown in \figref{fig:ICB-ECB}, demonstrate that the
$\mathcal{F}=0$ curve does not change significantly for a more
anisotropic slow wave distribution. Therefore, the picture of the
solar wind compressible fluctuations comprised of slow wave
fluctuations that are more anisotropically distributed than critical
balance is consistent with the observational data.

One of the limitations of this investigation is that the
Vlasov-Maxwell eigenfunctions are computed assuming isotropic
Maxellian equilibrium distributions.  Plasma measurements in the solar
wind frequently find anisotropic temperature distributions and
non-Maxwellian tails at high energy \citep{Marsch:1991,Marsch:2006}.
In particular, it is  possible that mirror modes may contribute to
the measured anti-correlation between the density and parallel
magnetic field, although an analysis of 10 years of \emph{Wind} data
show that only a small fraction of the solar wind measurements occupy
the region of parameter space near the mirror instability threshold
\citep{Bale:2009}. How the density-parallel magnetic
field correlation is altered by these conditions is an important issue
to be addressed in future work.

\section{Conclusion}
\label{sec:conc}
In this paper, we have presented the first application of kinetic
plasma theory to analyze and interpret the compressible fluctuations
in the inertial range of solar wind turbulence. This novel approach is
motivated by the fact that the dynamics in the solar wind plasma is
weakly collisional, a limit in which  MHD theory, used in all
previous related investigations, is formally invalid. Investigation
of the compressible fast and slow wave modes requires a kinetic
description to resolve both the wave dynamics and the collisionless
kinetic damping mechanisms. 
 
We identify quantitatively the linear kinetic wave modes of
Vlasov-Maxwell theory that correspond to the linear MHD fast and slow
wave modes, and verify that the general qualitative properties of the 
density-parallel magnetic field correlation $C(\delta n, \delta
B_{\parallel})$---that fast waves are positively correlated, and slow
waves are negatively correlated---remain unchanged for the weakly
collisional conditions of the solar wind plasma.

We then describe the procedure used to generate synthetic spacecraft
data used to interpret actual single-point spacecraft measurements of
turbulence in the solar wind.  We define and discuss the
\emph{quasi-linear premise}---that some properties of magnetized
plasma turbulence can be understood by modeling the turbulence as a
collection of randomly phased, linear waves---upon which the method of
synthetic spacecraft data is based. Theoretical arguments for the
validity of the quasi-linear premise are presented; a review of
supporting and conflicting evidence for the quasi-linear premise in
the literature is presented in \citet{Howes:2011f}. We outline how the
synthetic plasma data cubes are used to generate synthetic reduced
time series of single-point measurements that can be analyzed using
the same procedures as the actual spacecraft measurements.  Comparison
of the results of the analyses of both the synthetic and actual
spacecraft data enables novel physical interpretations of the solar
wind turbulence measurements \citep{Howes:2012a}.

Next, we use the synthetic spacecraft data method to predict the
characteristics of the density-parallel magnetic field correlation
$C(\delta n, \delta B_{\parallel})$ as a function of plasma $\beta_i$
in the presence of a turbulent spectrum of wave modes. Comparison of
these synthetic predictions of $C(\delta n, \delta B_{\parallel})$,
based on both linear MHD and linear Vlasov-Maxwell eigenfunctions, to
the results from an analysis of a 10-year data set of observations
from the \emph{Wind} spacecraft leads to the following conclusions:
\begin{enumerate}
\item The predicted $C(\delta n, \delta B_{\parallel})$ and its dependence on
plasma $\beta_i$ using linear MHD eigenfunctions is significantly
different from the prediction using linear Vlasov-Maxwell
eigenfunctions.  Only the prediction based on kinetic theory appears
to agree with the spacecraft measurements, leading to the expected
conclusion that MHD theory is inadequate to describe the compressible fluctuations
in the weakly collisional solar wind.
\item Strong \emph{a posteriori} evidence for the
validity of the quasi-linear premise is provided by the striking agreement between 
the observationally determined $C(\delta n, \delta B_{\parallel})$ over a 
very large statistical sample and the predicted
$C(\delta n, \delta B_{\parallel})$ based on synthetic spacecraft data.
\item The observed $C(\delta n, \delta B_{\parallel})$ computed 
in a companion work \citep{Howes:2012a} is consistent with a
statistically negligible kinetic fast wave energy contribution for the
large sample used in this study. Note, however, that the our companion
work also found that a very small fraction of the intervals have
$C(\delta n ,\delta B_{\parallel})>0$, possibly indicating a trace population of fast waves \citep{Howes:2012a}.
\item The quantitative dependence of $C(\delta n, \delta B_{\parallel})$ 
on the ion plasma beta $\beta_i$ provides evidence that the slow wave
fluctuations are not isotropically distributed, but rather have an
anisotropic distribution, that is possibly given by the condition of
critical balance \citep{Goldreich:1995} or that is more anisotropic
than critical balance.
\end{enumerate}

In conclusion, our analysis using kinetic theory to interpret the
compressible fluctuations motivates the following physical model of
the turbulent fluctuations in the solar wind inertial range. In this
model, the solar wind inertial range consists of a mixture of
turbulent fluctuations, with 90\% of the energy due to incompressible
\Alfvenic fluctuations and the remaining 10\% of the energy  due to
compressible slow wave fluctuations.  The \Alfvenic turbulent power is
distributed anisotropically in wave vector space according to critical
balance. The turbulent \Alfven wave dynamics advects and cascades the
slow wave fluctuations to smaller scales at the \Alfven wave
frequency. The slow wave turbulent power may be either critically
balanced, or more anisotropic than critical balance due to
collisionless damping of the slow wave fluctuations.

Since only the fast wave turbulent cascade is expected to nonlinearly
transfer energy to whistler waves at $k \rho_i \gtrsim 1$, and the
frequency mismatch between the fast and either the \Alfven or slow
waves should prevent non-linear coupling, there is little or no
transfer of large scale turbulent energy through the inertial range
down to whistler waves at small scales. Therefore, any whistler wave
fluctuations at scales $k \rho_i \gtrsim 1$ must be generated by some
other process, \emph{e.g.}, kinetic temperature anisotropy
instabilities \citep{Kasper:2002,Hellinger:2006,Bale:2009} or kinetic
drift instabilities driven by differential flow between protons and
alpha particles \citep{McKenzie:1993,Kasper:2008,Bourouaine:2011}.


Finally, the lack of statistically significant fast wave energy has
important implications for efficient numerical modeling of solar wind
turbulent fluctuations. This work demonstrates clearly the importance
of a kinetic approach to model adequately the turbulent fluctuations,
yet a general kinetic numerical treatment---\emph{e.g.}, the
Particle-In-Cell (PIC) method---in three spatial dimensions (required
for physically relevant modeling of the dominant nonlinear
interactions in solar wind turbulence \citep{Howes:2011a}) is too
computationally costly to be presently feasible. Fortunately, it is
possible to perform kinetic numerical simulations of solar wind
turbulence in three spatial dimensions using gyrokinetics, a rigorous,
low-frequency, anisotropic limit of kinetic theory
\citep{Rutherford:1968,Frieman:1982,Howes:2006,Schekochihin:2009}.
In the derivation of the gyrokinetic equation, the crucial step is an
averaging over the particle gyrophase, which leads to a theory with
the following properties: the fast/whistler wave and the cyclotron
resonances are discarded; all finite Larmor radius effects and
collisionless dissipation via the Landau resonance are retained; and
one of the dimensions of velocity in phase space is eliminated,
reducing the particle distribution function from six to five
dimensions. It has been previously pointed out that one cannot rule
out the contribution of fast wave or whistler wave physics to solar
wind turbulence, and that that therefore gyrokinetics is an incomplete
description of the turbulence \citep{Matthaeus:2008a}. The novel
observational analysis presented here suggests that the fast wave,
in fact, does not play a statistically significant role in the turbulent
dynamics of the inertial range, and that therefore a gyrokinetic
approach sufficiently describes all important physical mechanisms in
the solar wind inertial range.

%

\acknowledgments
K.~G.~K. and G.~G.~H. thank Jack Scudder for inspiring discussions.
The work has been supported by NSF CAREER Award AGS-1054061 and NASA
NNX10AC91G.

\appendix
\section{Normalized MHD Linear Eigenfunctions}
\label{app:MHDeig}
The ideal, compressible MHD equations are given by the continuity
equation,
\begin{equation}
\frac{\partial \rho}{\partial t} +\textbf{U}\cdot \nabla \rho=-\rho \nabla \cdot \textbf{U},
  \label{eqn:Continuity}
\end{equation}
the momentum equation,
\begin{equation}
\rho \left( \frac{\partial \textbf{U}}{\partial t} + \textbf{U}\cdot \nabla \textbf{U}\right)=
-\nabla \left( p + \frac{B^2}{8 \pi} \right) + \frac{(\textbf{B}\cdot
\nabla)\textbf{B}}{4 \pi}, \label{eqn:Momentum}
\end{equation}
the induction equation,
\begin{equation}
\frac{\partial \textbf{B}}{\partial t}=\nabla \times 
(\textbf{U} \times \textbf{B}),
  \label{eqn:Induction}
\end{equation}
and an adiabatic equation of state,
\begin{equation}
\frac{\partial p}{\partial t} +\textbf{U}\cdot \nabla p=-\gamma p \nabla \cdot \textbf{U}.
  \label{eqn:Pressure}
\end{equation}
Here the MHD fluid is completely described by its mass density $\rho=
n (m_i+m_e)$, fluid velocity $\V{U}$, magnetic field $\V{B}$, and
scalar thermal pressure $p$. The adiabatic index for a proton and
electron plasma is $\gamma=5/3$.  This system of equations  can be
derived rigorously from plasma kinetic theory in the MHD limit of
strong collisionality $\nu \gg \omega$, large scales compared to the
ion Larmor radius $k \rho_i \ll 1$, and non-relativistic conditions
\citep{Kulsrud:1983}.

Although the weakly collisional conditions of the solar wind plasma
violate the strong collisionality formally required for the validity
of MHD, this system has nevertheless been widely used to study the
turbulent dynamics of the solar wind inertial range.  That this
simplified approach has not met with widespread failure is likely due to the
fact that the turbulent dynamics in the inertial range is dominated by
\Alfvenic motions \citep{Belcher:1971,Tu:1995,Bruno:2005}. \Alfvenic fluctuations at inertial range scales are 
incompressible, and it has been shown that for anisotropic
fluctuations with $k_\parallel \ll k_\perp$ (the anisotropy generally
observed in the magnetized plasma turbulence
\citep{Robinson:1971,Zweben:1979,Montgomery:1981,Sahraoui:2010b,Chen:2011c}), the dynamics of 
\Alfvenic fluctuations are rigorously described by the equations of reduced MHD, 
even under weakly collisional conditions
\citep{Schekochihin:2009}. The compressible fast and slow wave
fluctuations, on the other hand, are substantially modified in weakly
collisional conditions, so their dynamics must be treated using
kinetic theory \citep{Kulsrud:1983,Schekochihin:2009}, as has been
demonstrated in the investigation presented here.

In this appendix, we derive the compressible MHD linear dispersion
relation and eigenfunctions in dimensionless units constructed
specifically for our study of compressible solar wind fluctutations.
Without loss of generality, we specify a wavevector $\V{k} = k_\perp
\xhat + k_\parallel \zhat$ and an equilibrium magnetic field $\V{B}_0
= B_0
\zhat$. We separate mean from fluctuating quantities using 
$\rho=\rho_0+\delta \rho$, $\V{U}=\delta \V{U}$, $\V{B}=\V{B}_0 + \delta
\V{B}$, and $p=p_0+\delta p$, where we specify that there is no mean
fluid velocity. We then linearize equations
(\ref{eqn:Continuity})--(\ref{eqn:Pressure}) and Fourier transform the
equations in both space and time. Next, we convert each of the
variables to dimensionless units and define the dimensionless frequency
and other  parameters as follows:
\begin{equation}
\begin{array}{ccccccccc}
\overline{\delta n} &=& \overline{\delta \rho} &=& \delta \rho/\rho_0 & \mbox{\hspace{1.0in}} & \overline{\omega} &=& \omega/(k v_A) \\

&&\overline{\delta \V{U}} &=& \delta \V{U}/v_A  & \mbox{\hspace{1.0in}} & \beta &=& c_s^2/v_A^2 \\
&&\overline{\delta \V{B}} &=& \delta \V{B}/B_0  & \mbox{\hspace{1.0in}} & c_s^2 &=& \gamma p_0/\rho_0 \\
&&\overline{\delta p} &=& \delta p/p_0 & \mbox{\hspace{1.0in}} & v_A^2 &=& B_0^2/4 \pi \rho_0\\ 
\end{array}
\end{equation}  
In this normalization, the linear dispersion relation takes the form
\begin{equation}
\left(\overline{\omega}^2-\cos^2\theta \right)
\left[ \overline{\omega}^4 - \overline{\omega}^2(1 +\beta) + \beta \cos^2 \theta
\right]=0
\end{equation}  
where the first factor in parentheses corresponds to the two \Alfven
wave solutions, and the second factor in the brackets corresponds to
the two slow and two fast wave solutions\footnote{Note that entropy
mode solution $\overline{\omega}=0$ has already been removed from this
dispersion relation since our focus here is on the propagating linear
wave modes.}. This demonstrates the important property that the
normalized compressible MHD linear dispersion relation depends on only
two parameters, $\omega/(k v_A) =\overline{\omega}_{MHD}( \beta,
\theta)$: the plasma beta $\beta$ and the angle $\theta$ between the
local mean magnetic field $\V{B}_0$ and the direction of the
wavevector $\V{k}$.

The linear \Alfven wave solutions have frequency
\begin{equation}
\overline{\omega}^2=\cos^2 \theta
\end{equation}  
and eigenfunctions specified in terms of  $\overline{\delta U_y}$ given by 
\begin{equation}
\overline{\delta B_y}=\pm \overline{\delta U_y}
\end{equation}  
\begin{equation}
\overline{\delta n}=\overline{\delta U_x} = 
\overline{\delta U_\parallel} = \overline{\delta B_x} = 
\overline{\delta B_\parallel}=\overline{\delta p}=0
\end{equation}  

The linear slow and fast wave solutions have frequency
\begin{equation}
\overline{\omega}^2=\frac{1+\beta \pm \sqrt{(1+\beta)^2-4 \beta \cos^2 \theta}}{2},
\label{eqn:freqFS}
\end{equation}  
where the plus sign corresponds to the fast wave, and the minus sign
to the slow wave.  The eigenfunctions, specified in terms of
$\overline{\delta B_\parallel}$, are given by
\begin{equation}
\overline{\delta n}= 
\frac{-\overline{\delta B_\parallel}\overline{\omega}^2}
{\overline{\omega}^2
-\beta\cos^2\theta}
\label{eqn:rhoFS}
\end{equation}
\begin{equation}
\overline{\delta  B_x}=-\overline{\delta B_\parallel} \cot \theta
\label{eqn:BperpFS}
\end{equation}
\begin{equation}
\overline{\delta U_x} = \overline{\omega} \overline{\delta B_\parallel} \csc \theta
\label{eqn:UperpFS}
\end{equation}
\begin{equation}
\overline{\delta U_\parallel} =
\frac{ \overline{\omega}\beta \cos \theta}
{\overline{\omega}^2
-\beta\cos^2\theta} \overline{\delta B_\parallel}
\label{eqn:UparFS}
\end{equation}
\begin{equation}
\overline{\delta U_y} =  \overline{\delta B_y} = 0
\end{equation}  

Since the synthetic data sets are created by specifying a spectrum of
fluctuations over some range of angles, the statistical correlations
of the fluctuations depend on only three factors: (a) angular power
distribution, (b) wave mode fraction, and (c) plasma $\beta$.

\section{Determining  Kinetic  Counterparts of Slow vs.~Fast MHD Waves}
\label{app:branch}

The kinetic fast and slow modes are both part of the double-valued
magnetosonic solution \citep{Krauss-Varban:1994} and are separated by
a branch cut from $\theta \approx 30 ^\circ$ to $0^\circ$ at $\beta_i
\approx 1.3$.  This branch cut is similar to the magnetic and acoustic
transition for the MHD fast and slow modes at $\beta=1$ for near
parallel wavevectors.  The kinetic fast and slow modes behave in an
analogous fashion: the near parallel kinetic fast mode has the same
phase velocity as the \Alfven mode for $\beta_i <1.3$; and the near
parallel kinetic slow mode matches the \Alfven mode for $\beta_i> 1.3$
(see the center column of \figref{fig:Polar}).  In this work, we label
wave modes using a scheme that identifies the modes at $\theta =
90^\circ$, where there is no ambiguity in identification, and then
follows the linear dispersion relation as $\theta$ is decreased to
zero.

The compressible inertial range linear wave modes with near parallel
wavevectors can be characterized as being either magnetic or acoustic.
The magnetic modes have phase velocities similar to the \Alfven wave,
while the acoustic modes are more heavily damped than either the
\Alfven or the magnetic modes. From this point of view, the fast mode
is the magnetic mode and the slow mode is the acoustic mode for
$\beta_i <1.3$, while the converse is true for $\beta_i >1.3$.
Interestingly, for $\beta_i < 1.3$, the magnetic (fast) mode 
has   $C(\delta n, \delta B_\parallel) \simeq 1$  and the acoustic (slow) mode
has  $C(\delta n, \delta B_\parallel) \simeq -1$; but, for
$\beta_i > 1.3$, both modes have $C(\delta n, \delta B_\parallel)
\simeq 0$ (see \figref{fig:proxy}). It is this transition that is responsible
for some of the qualitative features in the lower two rows of
\figref{fig:ssd_nbpar}. In particular, the abrupt jump on 
the $\mathcal{F}=1$ curve (black) at $\beta_i \sim 1.3$ in the left
and right columns (where fast waves are isotropically distributed and
so include this region of wavevector space) and in the $\mathcal{F}=0$
curve (orange) at $\beta_i \sim 1.3$ in the left column (where the
slow waves are isotropically distributed) are due to this
transition. 

This identification of modes has important physical ramifications,
especially at scales near the transition to the dissipation range.  As
length scales decrease towards the ion inertial length, it is the magnetic
mode, and not strictly the kinetic fast mode, that transitions to the
parallel whistler wave.  However, the results presented in this paper
suggest a lack of parallel slow wave energy as well as statistically
negligible fast wave energy in the solar wind inertial range, so the
difficulties of mode identification in this region of parameter space
may not be problematic for studies of  solar wind turbulence.



\end{document}